\documentclass[%
superscriptaddress,
twocolumn,
10pt,
aps,
prb,
floatfix,
]{revtex4-1}

\usepackage{amsfonts}
\usepackage{amsmath}
\usepackage{amssymb}
\usepackage{graphicx}
\usepackage{graphicx}
\usepackage{hyperref} \hypersetup{colorlinks=true,citecolor=blue,linkcolor=blue,urlcolor=blue}
\usepackage{xcolor,soul}

\makeatletter \renewcommand\frontmatter@abstractwidth{\dimexpr\textwidth\relax} \makeatother 

\setstcolor{red} 
\bibpunct{[}{]}{,}{n}{}{} 

\DeclareMathOperator*{\avg}{avg}
\DeclareMathOperator*{\std}{std}
\DeclareMathOperator*{\fraction}{frac}

\def\AFLOW{{\small AFLOW}}
\def\ICSD{{\small ICSD}}

\def\citeAFLOW{\cite{aflowPAPER,aflowBZ,curtarolo:art110,curtarolo:art63,curtarolo:art57,curtarolo:art53,curtarolo:art49,monsterPGM,aflowANRL,aflowPI}}
\def\citeAFLOWLIB{\cite{aflowlibPAPER,curtarolo:art92,curtarolo:art104,aflux}}

\begin{document}
\title{Machine learning modeling of superconducting critical temperature}
\author{Valentin Stanev}
\affiliation{Department of Materials Science and Engineering,
University of Maryland, College Park, MD 20742-4111, USA}
\affiliation{Center for Nanophysics and Advanced Materials, University of Maryland, College Park, Maryland 20742, USA}
\author{Corey Oses}
\affiliation{Department of Mechanical Engineering and Materials Science,
Duke University, Durham, North Carolina 27708, United States}
\affiliation{Center for Materials Genomics, Duke University, Durham, North Carolina 27708, United States}
\author{A. Gilad Kusne}
\affiliation{Department of Materials Science and Engineering,
University of Maryland, College Park, MD 20742-4111, USA}
\affiliation{National Institute of Standards and Technology, Gaithersburg, MD 20899, USA}
\author{Efrain Rodriguez}
\affiliation{Department of Chemistry and Biochemistry, University of Maryland, College Park, MD 20742, USA}
\affiliation{Center for Nanophysics and Advanced Materials, University of Maryland, College Park, Maryland 20742, USA}
\author{Johnpierre Paglione}
\affiliation{Department of Physics, University of Maryland, College Park, Maryland 20742, USA}
\affiliation{Center for Nanophysics and Advanced Materials, University of Maryland, College Park, Maryland 20742, USA}
\author{Stefano Curtarolo}
\affiliation{Department of Mechanical Engineering and Materials Science,
Duke University, Durham, North Carolina 27708, United States}
\affiliation{Center for Materials Genomics, Duke University, Durham, North Carolina 27708, United States}
\affiliation{Fritz-Haber-Institut der Max-Planck-Gesellschaft, 14195 Berlin-Dahlem, Germany}
\author{Ichiro Takeuchi}
\affiliation{Department of Materials Science and Engineering, University of Maryland, College Park, MD 20742-4111, USA}
\affiliation{Center for Nanophysics and Advanced Materials, University of Maryland, College Park, Maryland 20742, USA}

\date{\today}

\begin{abstract}
\noindent Superconductivity has been the focus of enormous research effort since its discovery more than a century ago.
Yet, some features of this unique phenomenon remain poorly understood; prime among these is the connection
between superconductivity and chemical/structural properties of materials.
To bridge the gap, several machine learning schemes are developed herein
to model the critical temperatures $\left(T_{\mathrm{c}}\right)$ of the
$12,000+$ known superconductors available via the SuperCon database.
Materials are first divided into two classes based on their $T_{\mathrm{c}}$ values,
above and below $10$~K,
and a classification model predicting this label is trained.
The model uses coarse-grained features based only on the chemical compositions.
It shows strong predictive power, with out-of-sample accuracy of about $92\%$.
Separate regression models 
are developed to predict the values of $T_{\mathrm{c}}$ for cuprate, iron-based, and ``low-$T_{\mathrm{c}}$'' compounds. 
These models also demonstrate good performance,
with learned predictors
offering potential insights into the mechanisms behind superconductivity in
different families of materials.
To improve the accuracy and interpretability of these models,
new features are incorporated using materials data
from the \AFLOW\ Online Repositories.
Finally, the classification and regression models are combined into a single integrated pipeline
and employed to search the entire Inorganic Crystallographic Structure Database (\ICSD) for potential new superconductors.
We identify more than 30 non-cuprate and non-iron-based oxides as candidate materials.
\end{abstract}

\maketitle

\section*{Introduction}

Superconductivity, despite being the subject of intense physics,
chemistry and materials science research for more than a century,
remains among one of the most puzzling scientific topics~\cite{SCSpecial_PSCC_2015}.
It is an intrinsically quantum phenomenon caused by a finite attraction between paired electrons,
with unique properties including zero DC resistivity, Meissner and Josephson effects, and
with an ever-growing list of current and potential applications.
There is even a profound connection between phenomena in the 
superconducting state and the Higgs mechanism in particle physics~\cite{PWAnderson_PR_1963}.
However, understanding the relationship between superconductivity and materials'
chemistry and structure presents significant theoretical and experimental challenges.
In particular, despite focused research efforts in the last 30 years,
the mechanisms responsible for high-temperature superconductivity in
cuprate and iron-based families remain elusive~\cite{Chu_PSCC_2015,Paglione_NatPhys_2010}.

Recent developments, however, allow a different approach to investigate what ultimately determines the superconducting
critical temperatures $\left(T_{\mathrm{c}}\right)$ of materials.
Extensive databases covering various measured and calculated materials properties have been created over
the years~\cite{ICSD,aflowPAPER,cmr_repository,Saal_JOM_2013,APL_Mater_Jain2013}.
The shear quantity of accessible information also makes possible,
and even necessary,
the use of data-driven approaches, \textit{e.g.}, statistical and machine learning (ML)
methods~\cite{Agrawal_APLM_2016,Lookman_MatInf_2016,Jain_JMR_2016,Mueller_MLMS_2016}.
Such algorithms can be
developed/trained on the variables collected in these databases,
and employed to predict macroscopic properties
such as the melting temperatures of binary compounds~\cite{Seko_PRB_2014},
the likely crystal structure at a given composition~\cite{Balachandran_SR_2015},
band gap energies~\cite{Pilania_SR_2016,curtarolo:art124} and
density of states~\cite{Pilania_SR_2016} of certain classes of materials.

Taking advantage of this immense increase of readily accessible and potentially relevant information, we develop several
ML methods modeling $T_{\mathrm{c}}$  from
the complete list of reported (inorganic) superconductors~\cite{SuperCon}.
In their simplest form, these methods take as input a number of predictors
generated from the elemental composition of each material.
Models developed with these basic features are surprisingly accurate, despite
lacking information of relevant properties, such as space group, electronic structure,
and phonon energies.
To further improve the predictive power of the models,
as well as the ability to extract useful information out of them,
another set of features are constructed based on crystallographic and electronic information
taken from the \AFLOW\ Online Repositories~\citeAFLOWLIB.

Application of statistical methods in the context of superconductivity began in the early
eighties with simple clustering methods~\cite{Villars_PRB_1988,Rabe_PRB_1992}.
In particular, three ``golden'' descriptors confine the sixty known (at the time) superconductors with
$T_{\mathrm{c}} > 10$~K to three small islands in space:
the averaged valence-electron numbers, orbital radii differences, and metallic
electronegativity differences.
Conversely, about $600$ other superconductors with $T_{\mathrm{c}} < 10$~K appear randomly dispersed
in the same space.
These descriptors were selected heuristically due to their
success in classifying binary/ternary structures and predicting stable/metastable ternary quasicrystals.
Recently, an investigation stumbled on this clustering problem again by observing a
threshold $T_{\mathrm{c}}$ closer to $\log\left(T_{\mathrm{c}}^{\mathrm{thres}}\right)\approx1.3$
$\left(T_{\mathrm{c}}^{\mathrm{thres}}=20~\mathrm{K}\right)$~\cite{curtarolo:art94}.
Instead of a heuristic approach, random forests and simplex fragments were
leveraged on the structural/electronic properties
data from the \AFLOW\ Online Repositories to find the optimum clustering descriptors.
A classification model was developed showing good performance.
Separately, a sequential learning framework was evaluated on superconducting materials,
exposing the limitations of relying on random-guess (trial-and-error) approaches for
breakthrough discoveries~\cite{Ling_IMMI_2017}.
Subsequently, this study also highlights the impact machine learning can have
on this particular field.
In another early work, 
statistical methods were used to find correlations between normal state properties and $T_{\mathrm{c}}$ 
of the metallic elements in the first six rows of the periodic table~\cite{Hirsch_PRB_1997}.
Other contemporary work hones in on specific materials~\cite{Owolabi_JSNM_2015,Ziatdinov_NanoTech_2016}
and families of superconductors~\cite{Klintenberg_CMS_2013,Owolabi_APTA_2014} (see also Ref.~\cite{Norman_RPP_2016}).

Whereas previous investigations explored several hundred compounds at most,
this work considers more than $16,000$ different compositions. 
These are extracted from the SuperCon database, which contains an exhaustive
list of superconductors, including many closely-related materials varying only by small changes in stoichiometry (doping plays a significant role in optimizing $T_{\mathrm{c}}$).
The order-of-magnitude increase in training data
(\textit{i}) presents crucial subtleties in chemical composition among related compounds,
(\textit{ii}) affords family-specific modeling exposing different superconducting mechanisms, and
(\textit{iii}) enhances model performance overall.
It also enables the optimization of several model construction procedures.
Large sets of independent variables can be constructed and rigorously filtered
by predictive power (rather than selecting them by intuition alone).
These advances are crucial to uncovering insights into the
emergence/suppression of superconductivity with composition.

As a demonstration of the potential of ML methods in looking for novel superconductors, we combined and 
applied several models to search for candidates among the roughly
$110,000$ different compositions contained in the Inorganic Crystallographic Structure Database (\ICSD).
The framework highlights 35  compounds with predicted $T_{\mathrm{c}}$'s
above 20~K for experimental validation.
Of these, some exhibit interesting chemical and structural similarities to cuprate superconductors, demonstrating the ability of the ML models to identify meaningful patterns in the data.
In addition, most materials from the list share a peculiar feature in their electronic band structure:
one (or more) flat/nearly-flat bands just below the energy of the highest occupied electronic state.
The associated large peak in the density of states (infinitely large in the limit of truly flat bands)
can lead to strong electronic instability, and has been discussed recently as one possible way to
high-temperature superconductivity~\cite{Kopnin_PRB_2011,Peotta_NComm_2015}.

\section*{Results}
\noindent \textbf{Data and predictors.}
\phantomsection
\label{data_pred}
The success of any ML method ultimately depends on access to reliable and plentiful data.
Superconductivity data used in this work is extracted from the SuperCon database~\cite{SuperCon},
created and maintained by the Japanese National Institute for Materials Science.
It houses information such as the $T_{\mathrm{c}}$
and reporting journal publication for superconducting materials known from experiment.
Assembled within it is a uniquely exhaustive list of all reported superconductors, 
as well as related non-superconducting compounds.

\begin{figure*} 
\centering
\includegraphics[width=\linewidth]{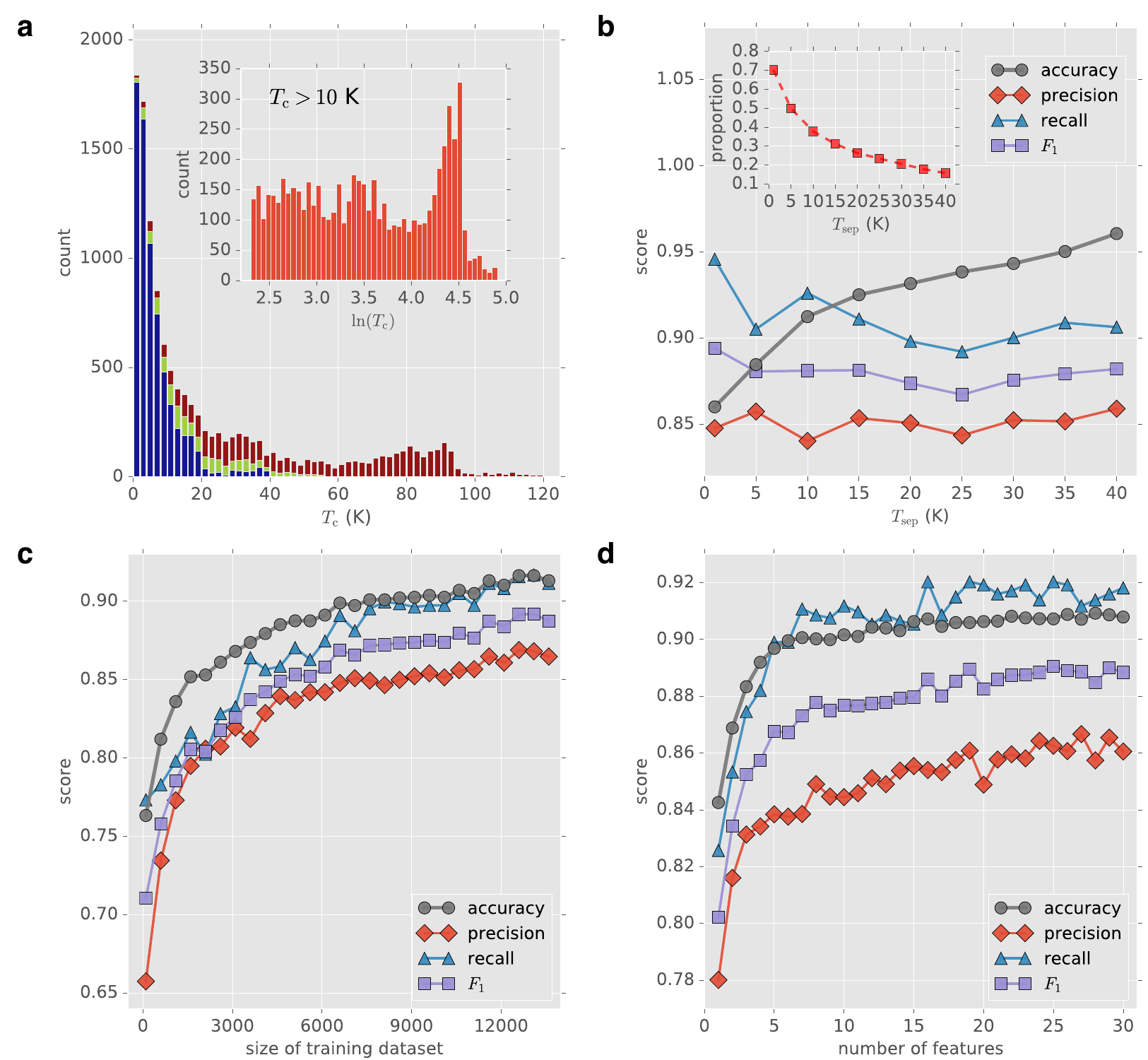}
\caption{\textbf{SuperCon dataset and classification model performance.}
(\textbf{a}) Histogram of materials categorized by
$T_{\mathrm{c}}$ (bin size is $2$~K, only those with finite $T_{\mathrm{c}}$ are counted).
Blue, green, and red denote ``low-$T_{\mathrm{c}}$'', iron-based, and cuprate superconductors, respectively.
In the inset: histogram of materials categorized by $\ln{(T_{\mathrm{c}})}$
restricted to those with $T_{\mathrm{c}} >10$~K.
(\textbf{b}) Performance of different classification models as a function of the threshold temperature 
$\left(T_{\mathrm{sep}}\right)$ that separates materials in two classes by $T_{\mathrm{c}}$.
Performance is measured by accuracy (gray), precision (red), recall (blue), and $F_{\mathrm{1}}$ score (purple).
The scores are calculated from predictions on an independent test set, \textit{i.e.}, one separate
from the dataset used to train the model.
In the inset: the dashed red curve gives the proportion of materials in the above-$T_{\mathrm{sep}}$ set.
(\textbf{c}) Accuracy, precision, recall, and $F_{\mathrm{1}}$
score as a function of the size of the training set with a fixed test set.
(\textbf{d}) Accuracy, precision, recall, and $F_{\mathrm{1}}$ as a function of the number of
predictors.}
\label{Class_score}
\end{figure*}

From SuperCon, we have extracted a list of approximately $16,400$ compounds, 
of which $4,000$ have no $T_{\mathrm{c}}$ reported (see Methods for details).
Of these, roughly $5,700$ compounds are cuprates and $1,500$ are iron-based
(about 35\% and 9\%, respectively), reflecting the significant research efforts invested in these two families.
The remaining set of about $8,000$ is a mix of various materials, including conventional phonon-driven superconductors
(\textit{e.g.}, elemental superconductor, A15 compounds), known unconventional superconductors like the
layered nitrides and heavy fermions, and many materials for which the mechanism of superconductivity
is still under debate (such as bismuthates and borocarbides).
The distribution of materials by $T_{\mathrm{c}}$ for the three groups is shown in Figure~\ref{Class_score}\textbf{a}.

Use of this data for the purpose of creating ML models can be problematic.
Training a model only on superconductors can lead to significant selection bias
that may render it ineffective when applied to new
materials~\footnote{\textit{N.B.}, a model suffering from selection bias
can still provide valuable statistical information about known superconductors.}.
Even if the model learns to correctly recognize factors promoting superconductivity,
it may miss effects that strongly inhibit it.
To mitigate the effect, we incorporate about $300$ materials found by H.
Hosono's group not to display superconductivity~\cite{Hosono_STAM_2015}.
However, the presence of non-superconducting materials, 
along with those without $T_{\mathrm{c}}$ reported in SuperCon, leads to a conceptual problem.
Surely, some of these compounds emerge as non-superconducting ``end-members'' from
doping/pressure studies, indicating no superconducting transition was observed despite some efforts to find one.
However, a transition may still exist,
albeit at experimentally difficult to reach or altogether inaccessible temperatures
(for most practical purposes below $10$~mK)~\footnote{There are theoretical arguments for this --- according
to the Kohn-Luttinger theorem, a
superconducting instability should be present as $T \rightarrow 0$ in any fermionic metallic system
with Coulomb interactions~\cite{Kohn_PRL_1965}.}.~\nocite{Kohn_PRL_1965}
This presents a conundrum:
ignoring compounds with no reported $T_{\mathrm{c}}$ disregards a potentially important
part of the dataset, while assuming $T_{\mathrm{c}} = 0$~K prescribes an inadequate description
for (at least some of) these compounds.
To circumvent the problem,
materials are first partitioned in two groups by their $T_{\mathrm{c}}$,
above and below a threshold temperature $\left(T_{\mathrm{sep}}\right)$,
for the creation of a classification model.
Compounds with no reported critical temperature can be classified in the ``below-$T_{\mathrm{sep}}$'' group
without the need to specify a $T_{\mathrm{c}}$ value (or assume it is zero).

For most materials, the SuperCon database provides
only the chemical composition and $T_{\mathrm{c}}$.
To convert this information into meaningful features/predictors (used interchangeably),
we employ the
Materials Agnostic Platform for Informatics and Exploration (Magpie)~\cite{Ward_ML_GFA_NPGCompMat_2016}.
Magpie computes a set of attributes for each material, including elemental property 
statistics like the mean and the standard deviation of 22 different elemental properties
(\textit{e.g.}, period/group on the periodic table,
atomic number, atomic radii, melting temperature), as well as electronic structure attributes, such as the average
fraction of electrons from the $s$, $p$, $d$ and $f$ valence shells among all
elements present.

The application of Magpie predictors, though appearing to lack \textit{a priori} justification, 
expands upon past clustering approaches by Villars and Rabe~\cite{Villars_PRB_1988,Rabe_PRB_1992}.
They show that, in the space of a few judiciously chosen
heuristic predictors, materials separate and cluster according to their
crystal structure and even complex properties such as high-temperature 
ferroelectricity and superconductivity.
Similar to these features, Magpie predictors capture significant chemical information, which
plays a decisive role in determining
structural and physical properties of materials.

Despite the success of Magpie predictors in modeling materials properties~\cite{Ward_ML_GFA_NPGCompMat_2016},
interpreting their connection to superconductivity presents a serious challenge.
They do not encode (at least directly) many important properties, particularly those
pertinent to superconductivity.
Incorporating features
like lattice type and density of states would undoubtedly lead to significantly more powerful and interpretable models.
Since such information is not generally available in SuperCon,
we employ data from the \AFLOW\ Online Repositories~\citeAFLOWLIB.
The materials database houses nearly 170 million properties calculated with
the software package \AFLOW~\citeAFLOW.
It contains information for the vast majority of compounds in the \ICSD~\cite{ICSD}.
Although the \AFLOW\ Online Repositories contain calculated properties,
the DFT results have been extensively validated with 
\ICSD\ records~\cite{curtarolo:art94,curtarolo:art96,curtarolo:art112,curtarolo:art115,curtarolo:art120,curtarolo:art124}.

\begin{figure*} 
\centering
\includegraphics[width=\linewidth]{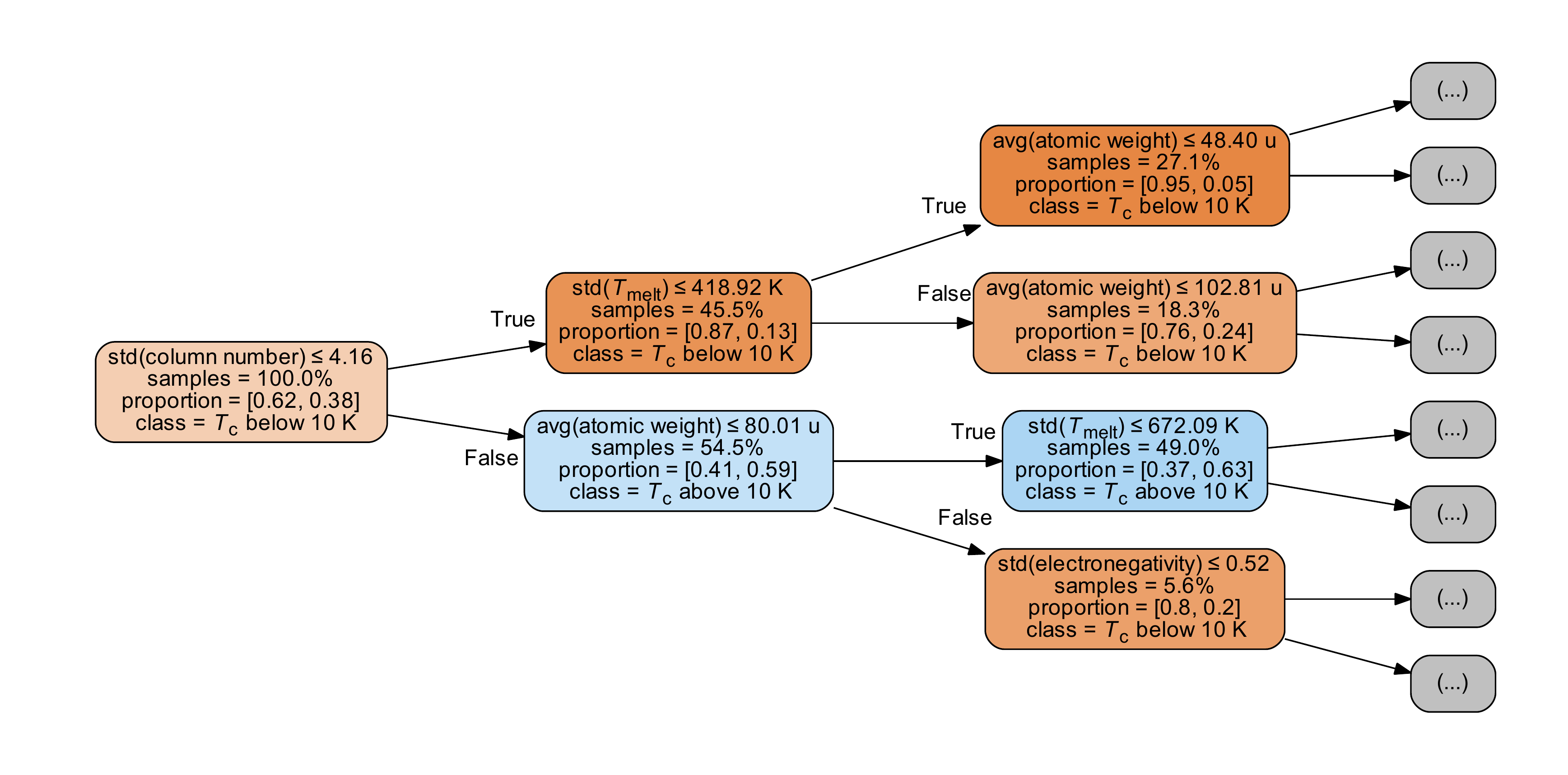}
\caption{\textbf{Schematic of the random forest ML approach.}
Example of a single decision tree used to classify materials depending on whether
$T_{\mathrm{c}}$ is above or below $10$~K.
A tree can have many levels, but only the three top are shown.
The decision rules leading to each subset are written inside individual rectangles.
The subset population percentage
is given by ``samples'', and the node color/shade
represents the degree of separation,
\textit{i.e.}, dark blue/orange illustrates a high proportion of
$T_{\mathrm{c}} >10$~K/$T_{\mathrm{c}} < 10$~K materials
(the exact value is given by ``proportion'').
A random forest consists of a large number --- could be hundreds or thousands --- of such individual trees.}
\label{tree_example}
\end{figure*}

Unfortunately, only a small subset of materials in SuperCon overlaps with those in the \ICSD:
about $800$ with finite $T_{\mathrm{c}}$ and less than $600$ are contained within \AFLOW.
For these, a set of 26 predictors are incorporated
from the \AFLOW\ Online Repositories, including structural/chemical information like the lattice type, space group,
volume of the unit cell, density, ratios of the lattice parameters,
Bader charges and volumes, and formation energy (see Supplementary Materials).
In addition, electronic properties are considered, including the
density of states near the Fermi level as calculated by \AFLOW.
Previous investigations exposed limitations in applying ML methods to a similar dataset
in isolation~\cite{curtarolo:art94}.
Instead, a framework is presented here for combining models built on Magpie descriptors
(large sampling, but features limited to compositional data) and \AFLOW\ features
(small sampling, but diverse and pertinent features).

Once we have a list of relevant predictors, various ML models can be applied to the
data~\cite{Bishop_ML_2006,Hastie_StatLearn_2001}.
All ML algorithms in this work are
variants of the random forest method~\cite{randomforests}.
Fundamentally, this approach combines many individual decision trees, where
each tree is a non-parametric supervised learning method used for 
modeling either categorical or numerical variables (\textit{i.e.},
classification or regression modeling).
A tree predicts the value of a target variable by learning simple decision rules
inferred from the available features (see Figure~\ref{tree_example} for an example).

Random forest is one of the most powerful, versatile, and widely-used ML methods~\cite{Caruana_2006}.
There are several advantages that make it especially suitable for this problem.
First, it can learn complicated non-linear dependencies from the data.
Unlike many other methods (\textit{e.g.}, linear regression),
it does not make any assumptions about the relationship between the predictors and the target variable.
Second, random forests are quite tolerant to heterogeneity in the training data.
It can handle both numerical and categorical data which, furthermore, does not
need extensive and potentially dangerous preprocessing, such as scaling or normalization.
Even the presence of strongly correlated predictors is not a problem for model
construction (unlike many other ML algorithms).
Another significant advantage of this method is that, by combining information from
individual trees, it can estimate the importance of each predictor, thus making the model more interpretable.
However, unlike model construction, determination of predictor importance is complicated by the presence of
correlated features.
To avoid this, standard feature selection procedures are employed along with
a rigorous predictor elimination scheme (based on their strength and correlation with others).
Overall, these methods
reduce the complexity of the models and improve our
ability to interpret them.

\noindent \textbf{Classification models.}
\phantomsection
\label{sec_class}
As a first step in applying ML methods to the dataset, a sequence of classification models
are created, each designed to separate materials into two distinct groups depending on whether
$T_{\mathrm{c}}$ is above or below some predetermined value.
The temperature that separates the two groups ($T_{\mathrm{sep}}$)
is treated as an adjustable parameter of the model, though some physical
considerations should guide its choice as well.
Classification ultimately allows compounds with no reported $T_{\mathrm{c}}$ to be used
in the training set by including them in the below-$T_{\mathrm{sep}}$ bin.
Although discretizing continuous variables is not generally recommended, in this case
the benefits of including compounds without $T_{\mathrm{c}}$ outweigh
the potential information loss.

In order to choose the optimal value of $T_{\mathrm{sep}}$, a series of random forest models
are trained with different threshold temperatures separating the two classes.
Since setting $T_{\mathrm{sep}}$ too low or too high creates strongly imbalanced classes
(with many more instances in one group), it is important to compare the models using several different metrics.
Focusing only on the accuracy (count of correctly-classified instances)
can lead to deceptive results.
Hypothetically, if $95\%$ of the observations in the dataset are in the below-$T_{\mathrm{sep}}$ group,
simply classifying all materials as such would
yield a high accuracy ($95\%$), while being trivial in any other sense.
To avoid this potential pitfall, three other standard metrics
for classification are considered: precision, recall, and $F_{\mathrm{1}}$ score.
They are defined using the values $tp$, $tn$, $fp$, and $fn$ for
the count of true/false positive/negative
predictions of the model:
\begin{eqnarray}
\text{accuracy} \equiv \frac{tp+tn}{tp+tn+fp+fn},
\label{accur}
\end{eqnarray}
\begin{eqnarray}
\text{precision}\equiv\frac{tp}{tp+fp},
\label{precision}
\end{eqnarray}
\begin{eqnarray}
\text{recall} \equiv\frac{tp}{tp+fn},
\label{recall}
\end{eqnarray}
\begin{eqnarray}
F_{\mathrm{1}}\equiv2*\frac{\text{precision}*\text{recall}}{\text{precision}+\text{recall}},
\label{f1}
\end{eqnarray}
where positive/negative refers to above-$T_{\mathrm{sep}}$/below-$T_{\mathrm{sep}}$.
The accuracy of a classifier is the total proportion of correctly-classified materials,
while precision measures the proportion of correctly-classified
above-$T_{\mathrm{sep}}$ superconductors out of all predicted above-$T_{\mathrm{sep}}$.
The recall is the proportion of correctly-classified above-$T_{\mathrm{sep}}$
materials out of all truly above-$T_{\mathrm{sep}}$ compounds.
While the precision measures the probability that a
material selected by the model actually has $T_{\mathrm{c}} > T_{\mathrm{sep}}$,
the recall reports how sensitive the model is to above-$T_{\mathrm{sep}}$ materials. 
Maximizing the precision or recall would require some compromise with
the other, \textit{i.e.}, a model that labels all materials as above-$T_{\mathrm{sep}}$ would have perfect recall but dismal precision.
To quantify the trade-off between recall and precision, their harmonic mean ($F_{\mathrm{1}}$ score) is
widely used to measure the performance of a classification model.
With the exception of accuracy, these metrics are not symmetric with respect to the exchange of positive and negative labels.

For a realistic estimate of the performance of each model,
the dataset is randomly split ($85\%/15\%$) into training and test subsets.
The training set is employed to fit the model, which is then applied to the test set for subsequent benchmarking.
The aforementioned metrics (Equations~\ref{accur}-\ref{f1}) calculated on the test set provide
an unbiased estimate of how well the model is expected to generalize to a new (but similar) dataset.
With the random forest method, similar estimates can be obtained intrinsically at the training stage.
Since each tree is trained only on a bootstrapped subset of the data, 
the remaining subset can be used as an internal test set.
These two methods for quantifying model performance usually yield very similar results.

\begin{figure*} 
\centering
\includegraphics[width=\linewidth]{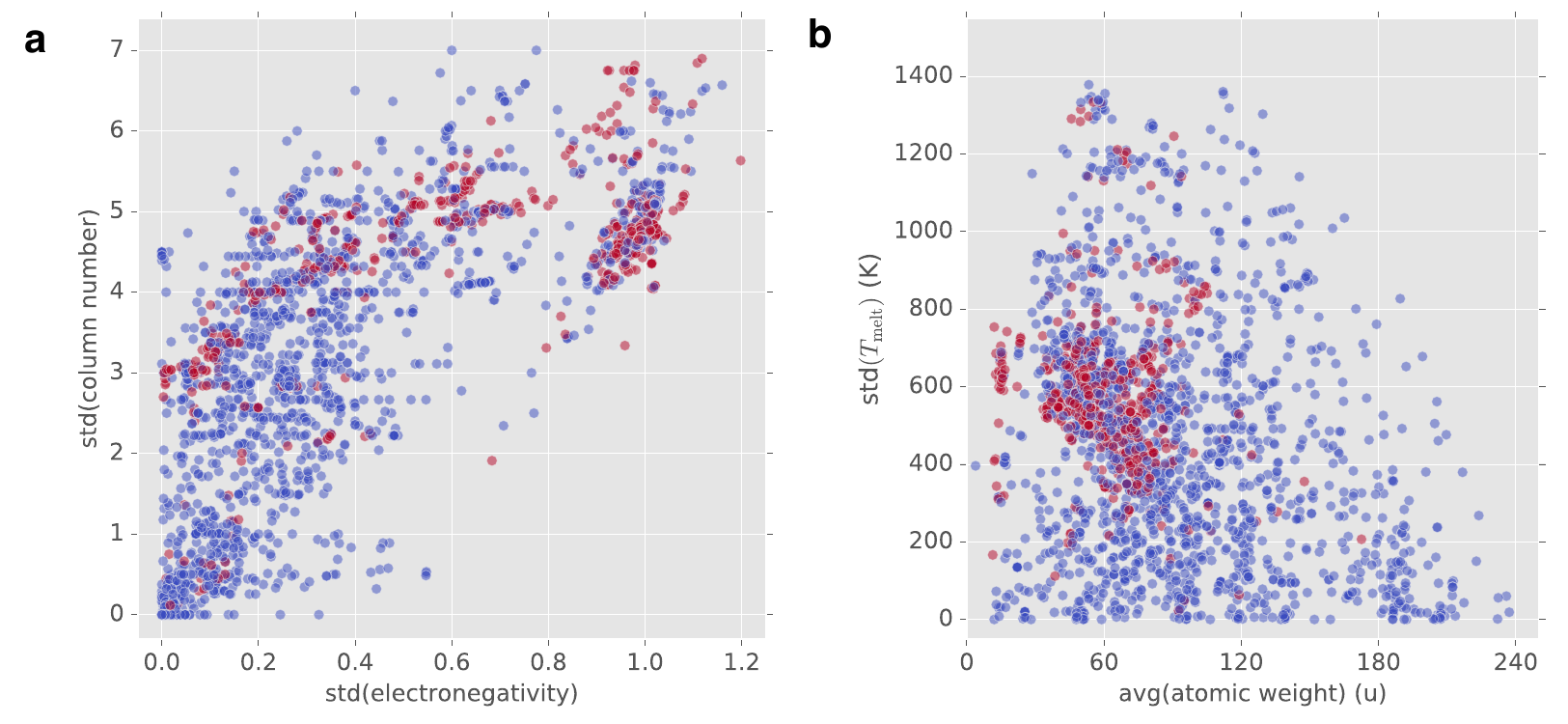}
\caption{\textbf{Scatter plots of $3,000$ superconductors in the space of the four most important classification predictors.}
Blue/red represent below-$T_{\mathrm{sep}}$/above-$T_{\mathrm{sep}}$ materials, where $T_{\mathrm{sep}} = 10$~K.
(\textbf{a}) Feature space of the first and second most important predictors:
standard deviations of the column numbers and electronegativities (calculated over the values for the constituent elements in each compound). 
(\textbf{b}) Feature space of the third and fourth most important predictors:
standard deviation of the elemental melting temperatures and average of the
atomic weights.}
\label{Class_features}
\end{figure*}

With the procedure in place, the models' metrics are evaluated for a range of $T_{\mathrm{sep}}$ and illustrated
in Figure~\ref{Class_score}\textbf{b}.
The accuracy increases as $T_{\mathrm{sep}}$ goes from $1$~K to $40$~K,
and the proportion of above-$T_{\mathrm{sep}}$ compounds drops from above $70\%$ to about $15\%$,
while the recall and $F_{\mathrm{1}}$ score generally decrease.
The region between $5-15$~K is especially appealing in (nearly) maximizing
all benchmarking metrics while balancing the sizes of the bins.
In fact, setting $T_{\mathrm{sep}}=10$~K is a particularly convenient choice.
It is also the temperature used in Refs~\cite{Villars_PRB_1988,Rabe_PRB_1992}
to separate the two classes, as it is just above the
highest $T_{\mathrm{c}}$ of all elements and pseudoelemental
materials (solid solution whose range of composition includes a pure element).
Here, the proportion of above-$T_{\mathrm{sep}}$ materials is approximately $38\%$ and
the accuracy is about $92\%$, \textit{i.e.}, the model can
correctly classify nine out of ten materials --- much better than random guessing.
The recall --- quantifying how well all above-$T_{\mathrm{sep}}$ compounds are labeled and,
thus, the most important metric when searching for new superconducting materials --- is even higher.
(Note that the models' metrics also depend on random factors such as the composition of the 
training and test sets, and their exact values can vary.)

\begin{table*} 
\centering
\caption{\textbf{The most relevant predictors and their importances for the classification and general regression models.}
$\avg(x)$ and $\std(x)$ denote the composition-weighted average and
standard deviation, respectively,
calculated over the vector of elemental values for each compound~\cite{Ward_ML_GFA_NPGCompMat_2016}.
For the classification model, all predictor importances are quite close.}
\begin{tabular}{ | c | l r| l r|}
\hline
predictor & 	\multicolumn{4}{|c|}{{\bf model}}
\\ 
\cline{2-5}
rank& \multicolumn{2}{|l|}{classification} & \multicolumn{2}{|l|}{regression (general; $T_{\mathrm{c}}>10$ K)} \\
\hline
1 & $\std ($column number$)$ & \setlength{\tabcolsep}{4pt} 0.26 & $\avg ($number of unfilled orbitals$)$ & \setlength{\tabcolsep}{4pt} 0.26\\
2 & $\std ($electronegativity$)$ & \setlength{\tabcolsep}{4pt} 0.26 & $\std ($ground state volume$)$ & \setlength{\tabcolsep}{4pt} 0.18\\
3 & $\std ($melting temperature$)$ & \setlength{\tabcolsep}{4pt} 0.23 & $\std ($space group number$)$ & \setlength{\tabcolsep}{4pt} 0.17\\
4 & $\avg ($atomic weight$)$ & \setlength{\tabcolsep}{4pt} 0.24 & $\avg ($number of $d$ unfilled orbitals$)$ & \setlength{\tabcolsep}{4pt} 0.17\\
5 & & \setlength{\tabcolsep}{4pt} - & $\std ($number of $d$ valence electrons$)$ & \setlength{\tabcolsep}{4pt} 0.12\\
6 & & \setlength{\tabcolsep}{4pt} - & $\avg ($melting temperature$)$ & \setlength{\tabcolsep}{4pt} 0.1\\
\hline
\end{tabular}
\label{Table1}
\end{table*}

The most important factors that determine the model's performance are the size of the available
dataset and the number of meaningful predictors.
As can be seen in Figure~\ref{Class_score}\textbf{c}, all
metrics improve significantly with the increase of the training set size. The effect is most dramatic for sizes between several hundred and few thousands instances, but there is no obvious saturation even for the largest available datasets. This validates efforts herein to incorporate as much relevant data as possible into model training.
The number of predictors is another very important model parameter.
In Figure~\ref{Class_score}\textbf{d},
the accuracy is calculated at each step of the backward feature elimination process.
It quickly saturates when the number of predictors reaches $10$.
In fact, a model with only 5 predictors achieves almost $90\%$ accuracy. (Note that these are the five most informative predictors, selected by the model out of the full list of 145 ones.)

For an understanding of what the model has learned, an analysis of the chosen predictors is needed.
In the random forest method, features can be ordered by their importance quantified via
the so-called Gini importance or
``mean decrease in impurity''~\cite{Bishop_ML_2006,Hastie_StatLearn_2001}.
For a given feature, it is the sum of the Gini impurity~\footnote{Gini impurity is calculated as
\unexpanded{$\sum_i p_i \left(1-p_i\right)$},
where \unexpanded{$p_i$} is the probability of randomly chosen data point 
from a given decision tree leaf to be in class 
\unexpanded{$i$}~\cite{Bishop_ML_2006,Hastie_StatLearn_2001}.}~\nocite{Bishop_ML_2006,Hastie_StatLearn_2001}
over the number of splits that include the feature, weighted by the number of samples
it splits, and averaged over the entire forest.
Due to the nature of the algorithm, the closer to the top of the tree a predictor is used,
the greater number of predictions it impacts.

Although correlations between predictors do not affect the model's ability to learn, it can distort importance estimates.
For example, a material property with a strong effect on $T_{\mathrm{c}}$ can be shared
among several correlated predictors.
Since the model can access the same information through any of these variables,
their relative importances are diluted across the group.
To reduce the effect and limit the list of predictors to a manageable size,
the backward feature elimination method is employed.
The process begins with a model constructed with the full list of predictors,
and iteratively removes the least significant one, rebuilding the model and recalculating importances
with every iteration.
(This iterative procedure is necessary since the ordering of the predictors by importance can change at each step.)
Predictors are removed until the accuracy drops
by no more than $2\%$, reducing the full list of 145 down to 5.
Furthermore, two of these predictors are strongly correlated with each other, and we remove the less important one. This
has a negligible impact on the model performance,
yielding four predictors total (see Table~\ref{Table1})
with an above $90\%$ accuracy score --- only slightly worse than the full model.
Scatter plots of the pairs of the most important predictors are shown in Figure~\ref{Class_features}, where
blue/red denotes whether the material is in the below-$T_{\mathrm{sep}}$/above-$T_{\mathrm{sep}}$ class.
Figure~\ref{Class_features}\textbf{a} shows a scatter plot of $3,000$ compounds 
in the space spanned by the standard deviations of the column numbers and electronegativities
calculated over the elemental values.
Superconductors with $T_{\mathrm{c}} > 10$~K tend to
cluster in the upper-right corner of the plot and in a relatively thin elongated region extending to the left of it.
In fact, the points in the upper-right corner represent mostly cuprate materials,
which with their complicated compositions and large number of elements are likely
to have high standard deviations in these variables.
Figure~\ref{Class_features}\textbf{b} shows
the same compounds projected in the space of the standard deviations of
the melting temperatures and the averages of the atomic weights of the elements forming each compound.
The above-$T_{\mathrm{sep}}$ materials tend to cluster in areas with lower mean atomic weights --- not
a surprising result given the role of phonons in conventional superconductivity.

For comparison, we create another classifier based on the average number of valence electrons,
metallic electronegativity differences, and orbital radii differences, \textit{i.e.}, the predictors used
in Refs.~\cite{Villars_PRB_1988,Rabe_PRB_1992} to cluster materials with $T_{\mathrm{c}} > 10$ K.
A classifier built only with these three predictors
is less accurate than both the full and the truncated models presented herein,
but comes quite close: the full model has about $3\%$ higher
accuracy and $F_{\mathrm{1}}$ score, while the truncated model with four predictors is less that $2\%$ more accurate.
The rather small (albeit not insignificant) differences demonstrates that even on the scale of the
entire SuperCon dataset, the predictors used by Villars and Rabe~\cite{Villars_PRB_1988,Rabe_PRB_1992}
capture much of the relevant chemical information for superconductivity.

\noindent \textbf{Regression models.}
After constructing a successful classification model, we now move to the more difficult challenge of predicting $T_{\mathrm{c}}$. 
Creating a regression model may enable better understanding of the factors controlling
$T_{\mathrm{c}}$ of known superconductors,
while also serving as an organic part of a system for identifying potential new ones.
Leveraging the same set of elemental predictors as the classification model, several regression models are presented
focusing on materials with $T_{\mathrm{c}} > 10$~K.
It avoids the problem of materials with no reported $T_{\mathrm{c}}$ with the assumption that,
if they were to exhibit superconductivity at all, their critical temperature would be below $10$~K.
Another problem is that the $T_{\mathrm{c}}$'s are unevenly distributed
over the $T_{\mathrm{c}}$ axis (see Figure~\ref{Class_score}\textbf{a}).
To avoid this,  $\ln{(T_{\mathrm{c}})}$ is used as the target variable instead of $T_{\mathrm{c}}$
(Figure~\ref{Class_score}\textbf{a} inset), which creates a more uniform distribution and 
is also considered a best practice when the range of a target
variable covers more than one order of magnitude (as in the case of $T_{\mathrm{c}}$).
Following this transformation, the dataset is parsed randomly ($85\%$/$15\%$) into training
and test subsets (similarly performed for the classification model).

\begin{figure*} 
\centering
\includegraphics[width=\linewidth]{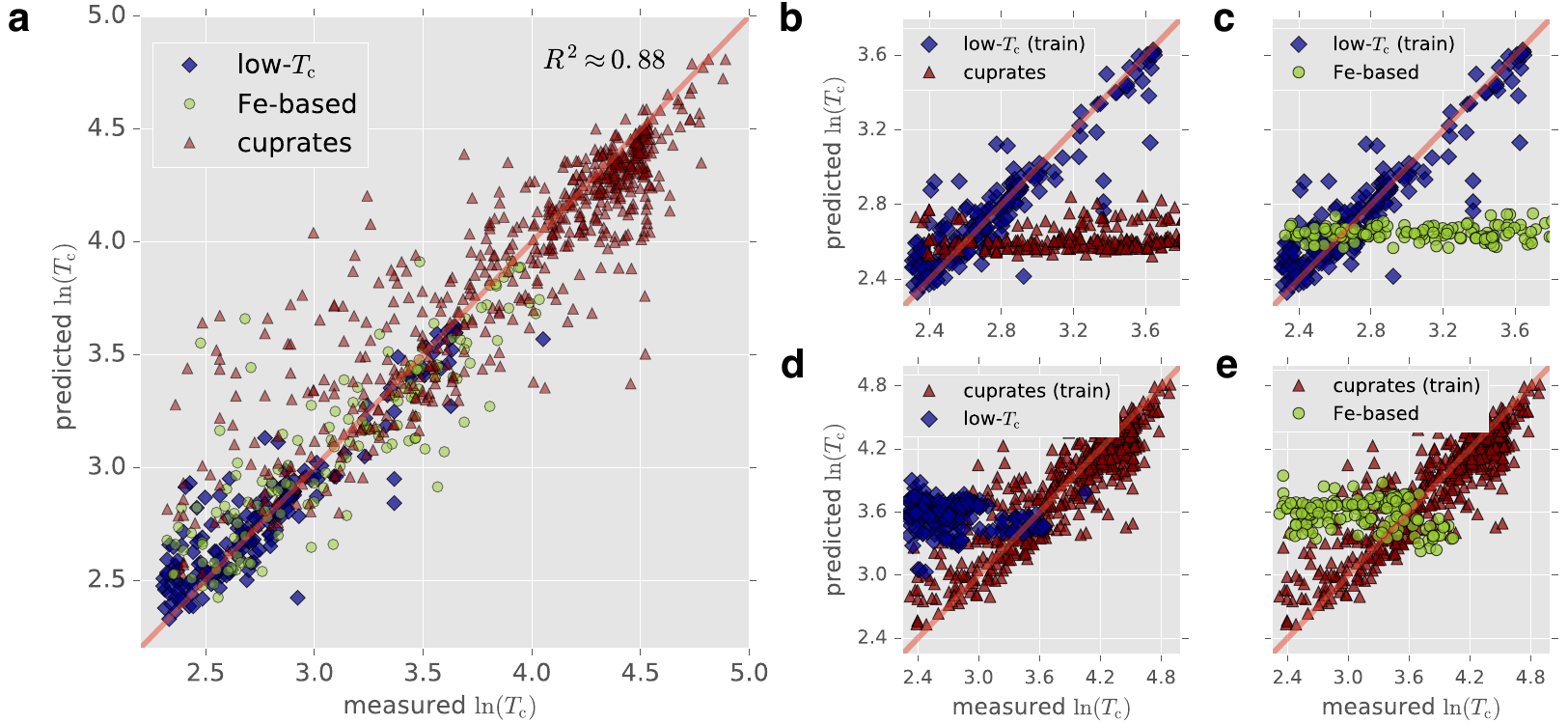}
\caption{\textbf{Benchmarking of regression models predicting $\ln(T_{\mathrm{c}})$.}
(\textbf{a}) Predicted 
\textit{vs.} measured $\ln(T_{\mathrm{c}})$ for the general regression model.
The test set comprises of a mix of low-$T_{\mathrm{c}}$, iron-based, and cuprate superconductors
with $T_{\mathrm{c}}>10$~K.
With an $R^2$ of about $0.88$, this one model can accurately predict
$T_{\mathrm{c}}$ for materials in different superconducting groups.
(\textbf{b and c}) Predictions of the regression model
trained solely on low-$T_{\mathrm{c}}$ compounds
for test sets containing cuprate and iron-based materials.
(\textbf{d and e}) Predictions of the regression model
trained solely on cuprates for test sets containing low-$T_{\mathrm{c}}$ and iron-based superconductors.
Models trained on a single group have no predictive power for materials from other groups.}
\label{Rerg_r2}
\end{figure*}

Present within the dataset are distinct families of superconductors with different driving
mechanisms for superconductivity, including cuprate and iron-based high-temperature superconductors,
with all others denoted ``low-$T_{\mathrm{c}}$'' for brevity (no specific mechanism in this group).
Surprisingly, a single regression model does reasonably well among the 
different families -- benchmarked on the test set, 
the model achieves $R^2 \approx 0.88$ (Figure~\ref{Rerg_r2}\textbf{a}).
It suggests that the random forest algorithm is flexible and powerful enough
to automatically separate the compounds into groups
and create group-specific branches with distinct predictors (no explicit group labels were used during training and testing).
As validation, three separate models are constructed trained only on a specific family, namely the
low-$T_{\mathrm{c}}$, cuprate, and iron-based superconductors, respectively.
Benchmarking on mixed-family test sets, the models performed well on compounds belonging
to their training set family while demonstrating no predictive power on the others.
Figures~\ref{Rerg_r2}\textbf{b}-\textbf{d} illustrate a cross-section of this comparison.
Specifically, the model trained on low-$T_{\mathrm{c}}$ compounds dramatically underestimates
the $T_{\mathrm{c}}$ of both high-temperature superconducting families (Figures~\ref{Rerg_r2}\textbf{b} and \textbf{c}),
even though this test set only contains compounds with $T_{\mathrm{c}} < 40$~K.
Conversely, the model trained on the cuprates tends to overestimate the $T_{\mathrm{c}}$
of low-$T_{\mathrm{c}}$ (Figure~\ref{Rerg_r2}\textbf{d}) and iron-based (Figure~\ref{Rerg_r2}\textbf{e}) superconductors.
This is a clear indication that superconductors from these groups have different factors determining their $T_{\mathrm{c}}$.
Interestingly, the family-specific models do not perform better than the general regression containing
all the data points: $R^2$ for the low-$T_{\mathrm{c}}$ materials is about $0.85$, for cuprates is just below $0.8$,
and for iron-based compounds is about $0.74$.
In fact, it is a purely geometric effect that
the combined model has the highest $R^2$.
Each group of superconductors contributes mostly to a distinct $T_{\mathrm{c}}$ range, and, as a result, the combined regression is better determined over longer temperature interval.

\begin{table*} 
\centering
\caption{\textbf{The most significant predictors and their importances for the three material-specific regression models.}
$\avg(x)$, $\std(x)$, $\max(x)$ and $\fraction(x)$ denote the composition-weighted average,
standard deviation, maximum, and fraction, respectively,
taken over the elemental values for each compound.
$l^2$-norm of a composition is calculated by $||x||_{2} = \sqrt{\sum_i x_i^2}$, where $x_i$ is the proportion of each element $i$ in the compound.}
\begin{tabular}{ | c | l r| l r |l r|}
\hline
pred. & 	\multicolumn{6}{|c|}{{\bf model}}
\\ 
\cline{2-7}
rank & \multicolumn{2}{|l|}{regression (low-$T_{\mathrm{c}}$)} & \multicolumn{2}{|l|}{regression (cuprates)}& \multicolumn{2}{|l|} {regression (Fe-based)} \\
\hline
1 & $\fraction (d$ valence electrons$)$ &\setlength{\tabcolsep}{4pt} 0.18 & $\avg ($number of unfilled orbitals$)$ &\setlength{\tabcolsep}{4pt} 0.22 & $\std ($column number$)$ &\setlength{\tabcolsep}{4pt} 0.17\\
2 & $\avg ($number of $d$ unfilled orbitals$)$ &\setlength{\tabcolsep}{4pt} 0.14 & $\std ($number of $d$ valence electrons$)$ &\setlength{\tabcolsep}{4pt} 0.13 & $\avg ($ionic character$)$ &\setlength{\tabcolsep}{4pt} 0.15\\
3 & $\avg ($number of valence electrons$)$ &\setlength{\tabcolsep}{4pt} 0.13 & $\fraction (d$ valence electrons$)$ &\setlength{\tabcolsep}{4pt} 0.13 & $\std ($Mendeleev number$)$ &\setlength{\tabcolsep}{4pt} 0.14\\
4 & $\fraction (s$ valence electrons$)$ &\setlength{\tabcolsep}{4pt} 0.11 & $\std ($ground state volume$)$ &\setlength{\tabcolsep}{4pt} 0.13 & $\std ($covalent radius$)$ &\setlength{\tabcolsep}{4pt} 0.14\\
5 & $\avg ($number of $d$ valence electrons$)$ &\setlength{\tabcolsep}{4pt} 0.09 & $\std ($number of valence electrons$)$ &\setlength{\tabcolsep}{4pt} 0.1 & $\max ($melting temperature$)$ &\setlength{\tabcolsep}{4pt} 0.14\\
6 & $\avg ($covalent radius$)$ &\setlength{\tabcolsep}{4pt} 0.09 & $\std ($row number$)$ &\setlength{\tabcolsep}{4pt} 0.08 & $\avg ($Mendeleev number$)$ &\setlength{\tabcolsep}{4pt} 0.14\\
7 & $\avg ($atomic weight$)$ &\setlength{\tabcolsep}{4pt} 0.08 & $||$composition$||_{2}$ &\setlength{\tabcolsep}{4pt} 0.07 & $||$composition$||_{2}$ &\setlength{\tabcolsep}{4pt} 0.11\\
8 & $\avg ($Mendeleev number$)$ &\setlength{\tabcolsep}{4pt} 0.07 & $\std ($number of $s$ valence electrons$)$ &\setlength{\tabcolsep}{4pt} 0.07 & &\setlength{\tabcolsep}{4pt} -\\
9 & $\avg ($space group number$)$ &\setlength{\tabcolsep}{4pt} 0.07 & $\std ($melting temperature$)$ &\setlength{\tabcolsep}{4pt} 0.07 & &\setlength{\tabcolsep}{4pt} -\\
10 & $\avg ($number of unfilled orbitals$)$ &\setlength{\tabcolsep}{4pt} 0.06 & &\setlength{\tabcolsep}{4pt} - & &\setlength{\tabcolsep}{4pt} -\\
\hline
\end{tabular}
\label{Table2}
\end{table*}

In order to reduce the number of predictors and increase the interpretability of these models without
significant detriment to their performance, a backward feature elimination process is again employed.
The procedure is very similar to the one described previously for the classification model,
with the only difference being that the reduction is guided by $R^2$ of the model, rather than the accuracy
(the procedure stops when $R^2$ drops by $3\%$).

\begin{figure*} 
\centering
\includegraphics[width=\linewidth]{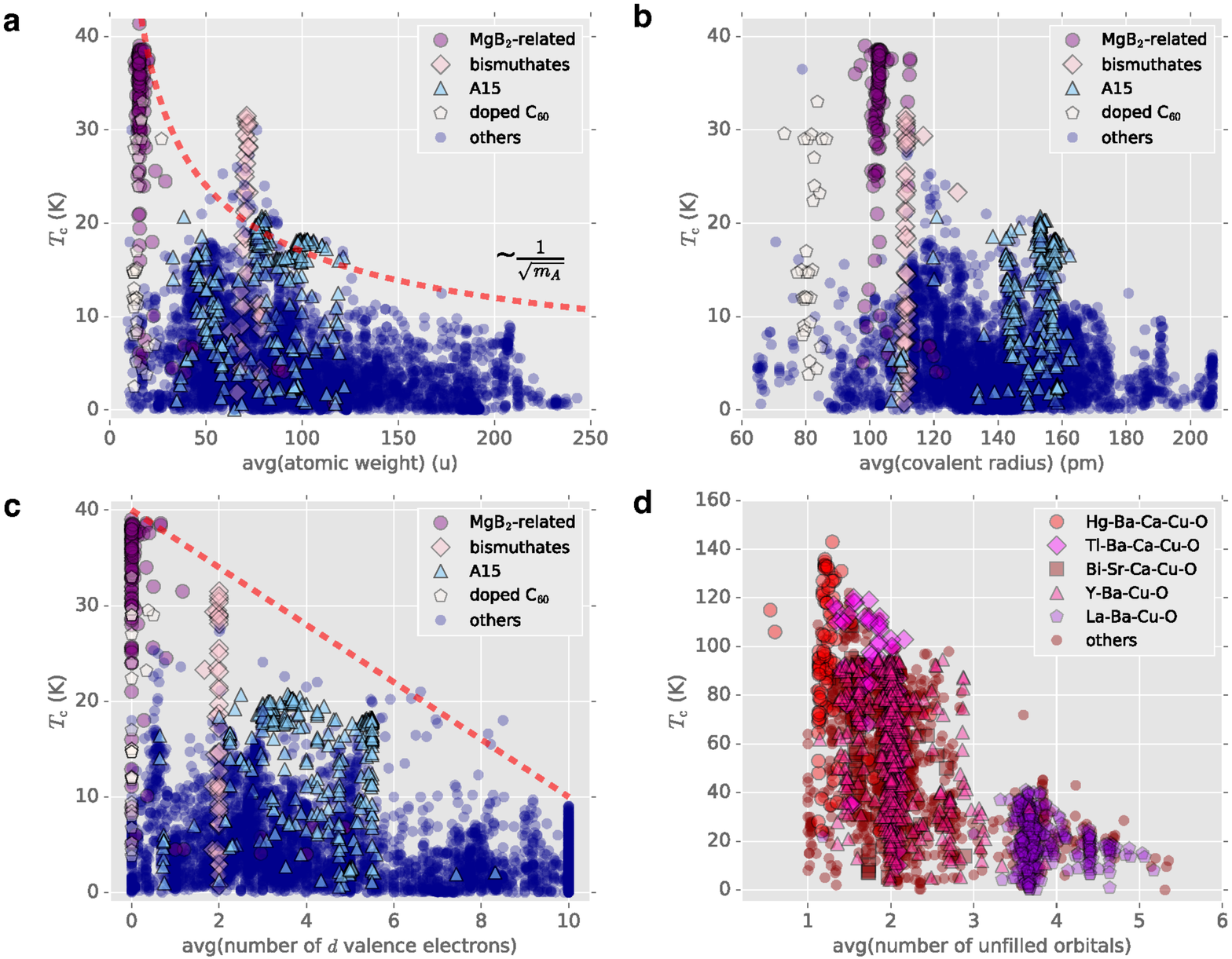}
\caption{\textbf{Scatter plots of $T_{\mathrm{c}}$ for superconducting materials in the space of significant,
family-specific regression predictors.}
For $4,000$ ``low-$T_{\mathrm{c}}$'' superconductors (\textit{i.e.}, non-cuprate and non-iron-based),
$T_{\mathrm{c}}$ is plotted
\textit{vs.} the
(\textbf{a}) average atomic weight,
(\textbf{b}) average covalent radius, and
(\textbf{c}) average number of $d$ valence electrons.
The dashed red line in (\textbf{a}) is $\sim 1/\sqrt{m_A}$.
Having low average atomic weight and low average number of $d$ valence
electrons are necessary (but not sufficient) conditions for achieving high $T_{\mathrm{c}}$
in this group.
(\textbf{d}) Scatter plot of $T_{\mathrm{c}}$ for all known superconducting cuprates \textit{vs.} the mean number of unfilled orbitals.
(\textbf{c and d}) suggest that the values of these predictors lead to
hard limits on the maximum achievable $T_{\mathrm{c}}$.}
\label{Tc_atomWeigth}
\end{figure*}

The most important predictors for the four models (one general and three family-specific) together with
their importances are shown in Tables~\ref{Table1} and \ref{Table2}.
Differences in important predictors across the family-specific models reflect the fact that
distinct mechanisms are responsible for driving superconductivity among these groups.
The list is longest for the low-$T_{\mathrm{c}}$ superconductors, reflecting the eclectic nature of
this group.
Similar to the general regression model,
different branches are likely created for distinct sub-groups.
Nevertheless, some important predictors have straightforward interpretation.
As illustrated in Figure~\ref{Tc_atomWeigth}\textbf{a},
low average atomic weight is a necessary (albeit not sufficient) condition for
achieving high $T_{\mathrm{c}}$ among the low-$T_{\mathrm{c}}$ group.
In fact, the maximum $T_{\mathrm{c}}$ for a given weight roughly follows $1/\sqrt{m_A}$.
Mass plays a significant role in conventional superconductors
through the Debye frequency of phonons, leading to the well-known formula $T_{\mathrm{c}} \sim 1/\sqrt{m}$,
where $m$ is the ionic mass.
Other factors like density of states are also important,
which explains the spread in $T_{\mathrm{c}}$ for a given $m_A$.
Outlier materials clearly lying above the $\sim 1/\sqrt{m_A}$ line include
bismuthates and chloronitrates, suggesting the conventional electron-phonon mechanism is not driving
superconductivity in these materials.
Indeed, chloronitrates exhibit a very weak isotope effect~\cite{Kasahara_PSCC_2015}, though
some unconventional electron-phonon coupling could still be relevant for superconductivity~\cite{Yin_PRX_2013}.
Another important feature for low-$T_{\mathrm{c}}$ materials 
is the average number of valence electrons.
This recovers the empirical relation first discovered by Matthias more than sixty years ago~\cite{Matthias_PR_1955}.
Such findings validate the ability of ML approaches
to discover meaningful patterns that encode true physical phenomena.

Similar $T_{\mathrm{c}}$-\textit{vs.}-predictor plots reveal more interesting and subtle features.
A narrow cluster of materials with $T_{\mathrm{c}} > 20$~K emerges in the context of the mean covalent radii of compounds --- another
important predictor for low-$T_{\mathrm{c}}$ superconductors.
The cluster includes (left-to-right) alkali-doped C$_{60}$, MgB$_2$-related compounds, and bismuthates.
The sector likely characterizes a region of strong covalent bonding and corresponding high-frequency phonon modes
that enhance $T_{\mathrm{c}}$ (however, frequencies that are too high become irrelevant for superconductivity).
Another interesting relation appears in the context of the average number of $d$ valence electrons. 
Figure~\ref{Tc_atomWeigth}\textbf{c} illustrates a fundamental bound on
$T_{\mathrm{c}}$ of all non-cuprate and non-iron-based superconductors.

A similar limit exists for cuprates based on the average number of unfilled orbitals (Figure ~\ref{Tc_atomWeigth}\textbf{d}). 
It appears to be quite rigid --- several data points found above it on inspection are actually 
incorrectly recorded entries in the database and were subsequently removed. 
The connection between $T_{\mathrm{c}}$ and the average number of unfilled orbitals~\footnote{The 
number of unfilled orbitals refers to the
electron configuration of the substituent elements before combining to form oxides.
For example, Cu has one unfilled orbital ([Ar]$4s^23d^9$) and Bi has 
three ([Xe]$4f^{14}6s^25d^{10}6p^3$). 
These values are averaged per formula unit.} 
may offer new insight into the mechanism for superconductivity in this family. 
Known trends include higher $T_{\mathrm{c}}$'s for structures that 
(\textit{i}) stabilize more than one superconducting Cu-O plane per unit cell 
and (\textit{ii}) add more polarizable cations such as Tl$^{3+}$ and Hg$^{2+}$ between these planes. 
The connection reflects these observations,
since more copper and oxygen per formula unit
leads to lower average number of unfilled orbitals (one for copper, two for oxygen). 
Further, the lower-$T_{\mathrm{c}}$ cuprates typically consist of Cu$^{2-}$/Cu$^{3-}$-containing 
layers stabilized by the addition/substition of hard cations, 
such as Ba$^{2+}$ and La$^{3+}$, respectively.
These cations have a large number of unfilled orbitals, thus increasing the compound's average. 
Therefore, the ability of between-sheet cations 
to contribute charge to the Cu-O planes may be indeed quite important.
The more polarizable the $A$ cation, the more electron density it can contribute 
to the already strongly covalent Cu$^{2+}$--O bond.

\noindent \textbf{Including \AFLOW.}
The models described previously demonstrate
surprising accuracy and predictive power, especially considering the difference between the
relevant energy scales of most Magpie predictors (typically in the range of eV) and superconductivity (meV scale).
This disparity, however, hinders the interpretability of the models,
\textit{i.e.}, the ability to extract meaningful physical correlations.
Thus, it is highly desirable to create accurate ML models with features based on
measurable macroscopic properties of the actual compounds
(\textit{e.g.}, crystallographic and electronic properties)
rather than composite elemental predictors.
Unfortunately, only a small subset of materials in SuperCon
is also included in the \ICSD:
about $1,500$ compounds in total, only about $800$ with finite $T_{\mathrm{c}}$,
and even fewer are characterized with {\it ab initio} calculations.
In fact, a good portion of known superconductors are disordered (off-stoichiometric) materials and
notoriously challenging to address with DFT calculations.
Currently, much faster and efficient methods are becoming available~\cite{curtarolo:art110}
for future applications.

To extract suitable features, data is incorporated from
the \AFLOW\ Online Repositories --- a database of 
DFT calculations managed by the software package \AFLOW.
It contains information for the vast majority of compounds 
in the \ICSD\ and about 550 superconducting materials.
In Ref.~\onlinecite{curtarolo:art94}, several 
ML models using a similar set of materials are presented.
Though a classifier shows good accuracy, attempts to create a 
regression model for $T_{\mathrm{c}}$ led to disappointing results.
We verify that using Magpie predictors for the superconducting compounds in the \ICSD\
also yields an unsatisfactory regression model.
The issue is not the lack of compounds \textit{per se}, as
models created with randomly drawn subsets from SuperCon with 
similar counts of compounds perform much better.
In fact, the
problem is the chemical sparsity of superconductors in the \ICSD, \textit{i.e.},
the dearth of closely-related compounds (usually created by chemical substitution).
This translates to compound scatter in predictor space --- a challenging learning environment for the model.

\begin{table} 
\centering
\caption{List of potential superconductors identified by the pipeline.  
Also shown are their ICSD numbers and symmetries. 
Note that for some compounds there are several entries.
All of the materials contain oxygen.}
\begin{tabular}{l c c }
\hline
compound & ICSD & SYM  \\
\hline
CsBe(AsO$_4$) & 074027 & orthorhombic \\
RbAsO$_2$ & 413150 & orthorhombic \\
KSbO$_2$ & 411214 & monoclinic \\
RbSbO$_2$ & 411216 & monoclinic \\
CsSbO$_2$ & 059329 & monoclinic \\
\hline
AgCrO$_2$ & 004149/025624 & hexagonal \\
K$_{0.8}$(Li$_{0.2}$Sn$_{0.76}$)O$_2$ & 262638 & hexagonal \\
\hline
Cs(MoZn)(O$_3$F$_3$)& 018082 & cubic \\
\hline
Na$_3$Cd$_2$(IrO$_6$) & 404507 & monoclinic \\
Sr$_3$Cd(PtO$_6$) & 280518 & hexagonal \\
Sr$_3$Zn(PtO$_6$) & 280519 & hexagonal \\
\hline
(Ba$_5$Br$_2)$Ru$_2$O$_9$ & 245668 & hexagonal \\
\hline
Ba$_4$(AgO$_2$)(AuO$_4)$ & 072329 & orthorhombic \\
Sr$_5$(AuO$_4$)$_2$ & 071965 & orthorhombic \\
\hline
RbSeO$_2$F & 078399 & cubic \\
CsSeO$_2$F & 078400 & cubic \\
KTeO$_2$F & 411068 & monoclinic \\
\hline
Na$_2$K$_4$(Tl$_2$O$_6$) & 074956 & monoclinic \\
\hline
Na$_3$Ni$_2$BiO$_6$ & 237391 & monoclinic \\
Na$_3$Ca$_2$BiO$_6$ & 240975 & orthorhombic\\
\hline

CsCd(BO$_3$) & 189199 & cubic \\
\hline
K$_2$Cd(SiO$_4)$ & 083229/086917 & orthorhombic \\
Rb$_2$Cd(SiO$_4$) & 093879 & orthorhombic \\
K$_2$Zn(SiO$_4$) & 083227 & orthorhombic \\
K$_2$Zn(Si$_2$O$_6$) & 079705 & orthorhombic \\
\hline

K$_2$Zn(GeO$_4$) & 069018/085006/085007 & orthorhombic \\
(K$_{0.6}$Na$_{1.4})$Zn(GeO$_4)$ & 069166 & orthorhombic \\
K$_2$Zn(Ge$_2$O$_6$) & 065740 & orthorhombic \\
Na$_6$Ca$_3$(Ge$_2$O$_6$)$_3$ & 067315 & hexagonal \\
Cs$_3$(AlGe$_2$O$_7$) & 412140 & monoclinic \\
K$_4$Ba(Ge$_3$O$_9$) & 100203 & monoclinic \\
K$_{16}$Sr$_4$(Ge$_3$O$_9$)$_{4}$ & 100202 & cubic \\
K$_3$Tb[Ge$_3$O$_8$(OH)$_2$] & 193585 & orthorhombic \\
K$_3$Eu[Ge$_3$O$_8$(OH)$_2$] & 262677 & orthorhombic \\
\hline
KBa$_6$Zn$_4$(Ga$_7$O$_{21}$) & 040856 & trigonal \\
\hline
\end{tabular}
\label{Table3}
\end{table}

The chemical sparsity in \ICSD\ superconductors is a significant hurdle, even when both sets of predictors
(\textit{i.e.}, Magpie and \AFLOW\ features) are combined via feature fusion.
Additionally, this approach alone neglects the majority of the $16,000$ compounds available via SuperCon.
Instead, we constructed separate models employing
Magpie and \AFLOW\ features, and then judiciously combined the results
to improve model metrics --- known as late or decision-level fusion.
Specifically, two independent classification models are developed,
one using the full SuperCon dataset and Magpie predictors, and another based on
superconductors in the \ICSD\ and \AFLOW\ predictors.
Such an approach can improve the recall, for example, in the case where we classify ``high-$T_{\mathrm{c}}$''
superconductors as those predicted by either model to be above-$T_{\mathrm{sep}}$.
Indeed, this is the case here where, separately, the models obtain a recall of $40\%$ and $ 66\%$, respectively, and
together achieve a recall of about $76\%$ (subject to small fluctuations due to variations in the test sets).
In this way, the models' predictions complement each other in a constructive way such that
above-$T_{\mathrm{sep}}$ materials missed by one model (but not the other) are now accurately classified.

\begin{figure*}
\centering
\includegraphics[width=\linewidth]{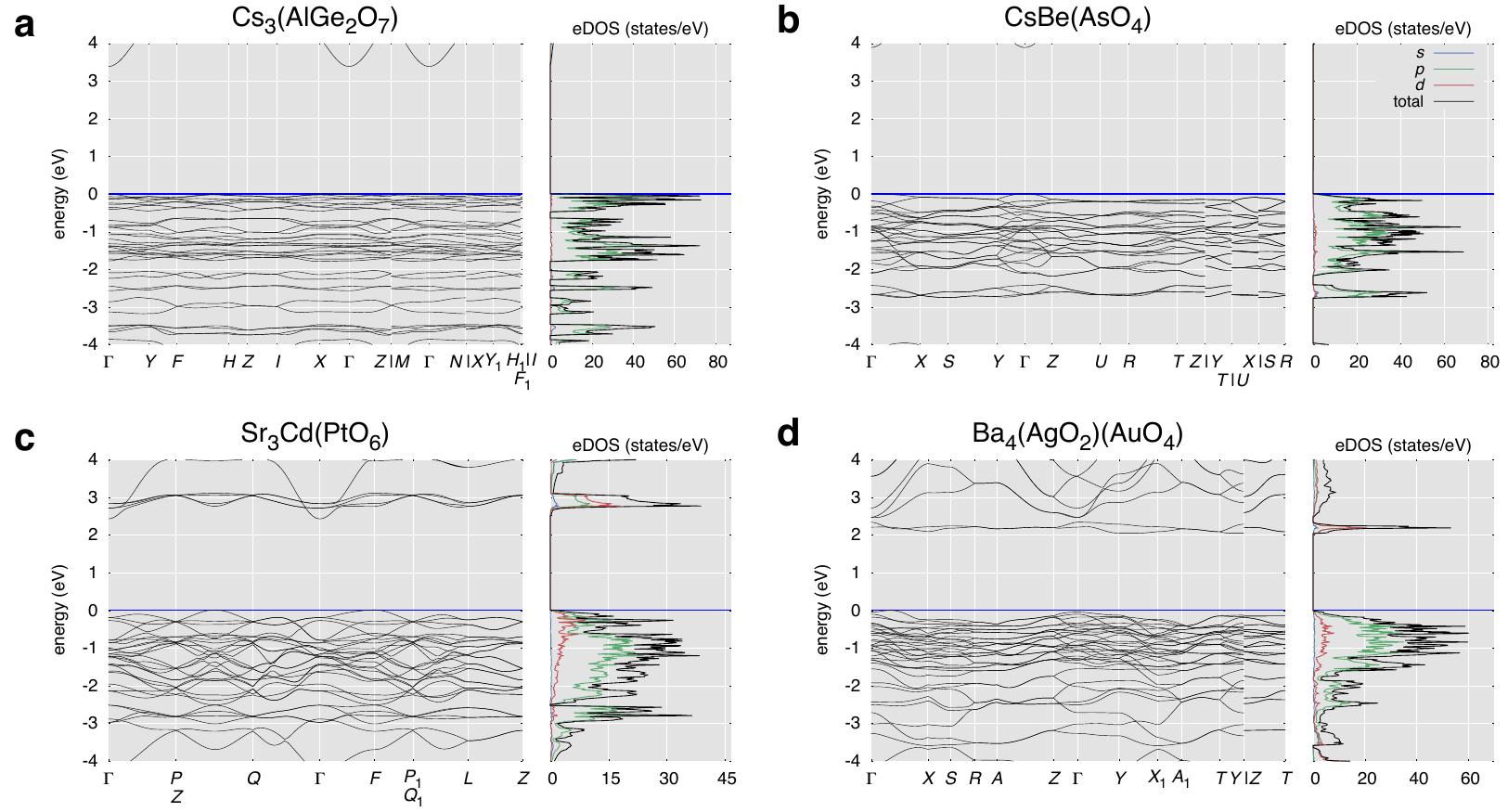}
\caption{\textbf{DOS of four compounds identified by the ML algorithm as potential materials with $T_{\mathrm{c}} > 20$~K.}
The partial DOS contributions from $s$, $p$ and $d$ electrons and total DOS are shown in blue, green, red, and black, respectively.
The large peak just below $E_F$ is a direct consequence of the flat band(s) present in all these materials.
These images were generated automatically via \AFLOW~\cite{curtarolo:art53}.
In the case of substantial overlap among \textbf{k}-point labels, the right-most label is offset below.}
\label{flat_bands}
\end{figure*}

\noindent \textbf{Searching for new superconductors in the \ICSD.}
As a final proof of concept demonstration,
the classification and regression models
described previously are integrated in one pipeline
and employed to screen the entire \ICSD\ database for candidate ``high-$T_{\mathrm{c}}$'' superconductors.
(Note that ``high-$T_{\mathrm{c}}$'' is a simple label,
the precise meaning of which can be adjusted.)
Similar tools power high-throughput screening workflows for materials with desired
thermal conductivity and magnetocaloric properties~\cite{curtarolo:art120,Bocarsly_ChemMat_2017}.
As a first step, the full set of Magpie predictors are generated for all
compounds in SuperCon.
A classification model similar to the one presented above is constructed,
but trained only on materials in SuperCon and not in the \ICSD\ (used
as an independent test set).
The model is then applied on the \ICSD\ set
to create a list of materials with predicted $T_{\mathrm{c}}$ above $10$~K.
Opportunities for model benchmarking are limited to those
materials both in the SuperCon and \ICSD\ datasets, though this test
set is shown to be problematic.
The set includes about $1,500$ compounds, though $T_{\mathrm{c}}$ is reported for only about half of them.
The model achieves an impressive accuracy of $0.98$, which is overshadowed by the fact that
$96.6\%$ of these compounds belong to the $T_{\mathrm{c}} < 10$~K class.
The precision, recall, and $F_{\mathrm{1}}$ scores are about $0.74$,
$0.66$, and $0.70$, respectively.
These metrics are lower than the estimates
calculated for the general classification model,
which is not unexpected given that this set cannot
be considered randomly selected.
Nevertheless, the performance suggests a good opportunity to identify new candidate superconductors.

Next in the pipeline, the list is fed into a random forest regression
model (trained on the entire SuperCon database)
to predict $T_{\mathrm{c}}$.
Filtering on the materials with $T_{\mathrm{c}} > 20$~K,
the list is further reduced to about $2,000$ compounds.
This count may appear daunting, but should
be compared with the total number of compounds in the database --- about $110,000$.
Thus, the method selects less than two percent of all materials,
which in the context of the training set (containing more than $20\%$ with ``high-$T_{\mathrm{c}}$''),
suggests that the model is not overly biased toward predicting high critical temperatures.

The vast majority of the compounds identified as
candidate superconductors are cuprates,
or at least compounds that contain copper and oxygen.
There are also some materials clearly related to the iron-based superconductors.
The remaining set has 35 members, and is composed of materials that are not obviously
connected to any high-temperature superconducting families (see Table~\ref{Table3})~\cite{Note5}.
None of them is predicted to have
$T_{\mathrm{c}}$ in excess of $40$~K, which is not surprising, given that no such instances exist in the training dataset. All contain oxygen --- also not a surprising result, since the group of
known superconductors with $T_{\mathrm{c}} > 20$~K is dominated by oxides.

The list comprises several distinct groups. Especially interesting are the compounds containing heavy metals (such as Au, Ir, Ru), metalloids (Se, Te), and heavier post-transition metals (Bi, Tl), which are or could be pushed into interesting/unstable oxidation states. Charge doping and/or pressure may be needed to drive these materials  into a superconducting state. The most surprising and non-intuitive of the compounds in the list are the silicates and the germanates. These materials form corner-sharing SiO$_4$ or GeO$_4$ polyhedra, not unlike quartz glass, and also have counter cations with full or empty shells such as Cd$_2$$^+$ or K$^+$.
Converting these insulators to metals (and possibly superconductors) likely requires 
significant charge doping. However, the similarity between these compounds and cuprates is meaningful.  In compounds like K$_2$CdSiO$_4$ or K$_2$ZnSiO$_4$,  K$_2$Cd (or K$_2$Zn) unit carries a 4+ charge that offsets the (SiO$_4$)$^{4-}$ (or (GeO$_4$)$^{4-}$) charges. This is reminiscent of the way Sr$_2$ balances the (CuO$_4$)$^{4-}$ unit in Sr$_2$CuO$_4$. 
Such chemical similarities based on charge balancing and stoichiometry were likely identified and exploited by the ML algorithms.

The electronic properties calculated by \AFLOW\ offer additional insight into the results of the search, and suggest a possible connection among these candidate.
Plotting the electronic structure of the potential superconductors exposes an extremely peculiar feature shared
by  almost all --- one or several (nearly) flat bands just below the energy of the highest occupied electronic state.
Such bands lead to a large peak in the DOS (see Figure~\ref{flat_bands}) and
can cause a significant enhancement in $T_{\mathrm{c}}$.
Peaks in the DOS elicited by van Hove singularities can enhance $T_{\mathrm{c}}$ 
if sufficiently close to $E_{\mathrm{F}}$~\cite{Labbe_PRL_1967,Hirsch_PRL_1986,Dzyaloshinskii_JETPLett_1987}.
However, note that unlike typical van Hove points, a true flat band creates divergence
in the DOS (as opposed to its derivatives), which in turn leads to a critical temperature
dependence linear in the pairing interaction strength, rather than the usual exponential relationship
yielding lower $T_{\mathrm{c}}$~\cite{Kopnin_PRB_2011}.
Additionally, there is significant similarity
with the band structure and DOS of layered
BiS$_2$-based superconductors~\cite{Yazici_PSCC_2015}.

This band structure feature came as the surprising
result of applying the ML model.
It was not sought for, and, moreover,
no explicit information about the electronic band structure has been
included in these predictors.
This is in contrast to the algorithm presented in Ref.~\onlinecite{Klintenberg_CMS_2013},
which was specifically designed to filter \ICSD\ compounds based on several preselected electronic structure features.

While at the moment it is not clear if some (or indeed any) of these compounds are really superconducting,
let alone with $T_{\mathrm{c}}$'s above 20~K,
the presence of this highly unusual electronic structure feature is encouraging.
Attempts to synthesize several of these compounds are already underway.

\section*{Discussion}
Herein, several machine learning tools are developed to study the critical temperature of superconductors.
Based on information from the SuperCon database, initial coarse-grained
chemical features are generated using the Magpie software.
As a first application of ML methods, materials are divided into two classes depending on
whether $T_{\mathrm{c}}$ is above or below $10$~K.
A non-parametric random forest classification model is constructed
to predict the class of superconductors.
The classifier shows excellent performance, with out-of-sample accuracy and $F_{\mathrm{1}}$
score of about $92\%$.
Next, 
several successful random forest regression models are created to predict the value of $T_{\mathrm{c}}$, 
including separate models for three material sub-groups, \textit{i.e.},
cuprate, iron-based, and ``low-$T_{\mathrm{c}}$'' compounds.
By studying the importance of predictors for each family of superconductors,
insights are obtained about the
physical mechanisms driving superconductivity among the different groups.
With the incorporation of crystallographic-/electronic-based features 
from the \AFLOW\ Online Repositories, the ML models are further improved. 
Finally, we combined these models into one integrated pipeline, which is employed to search the entire
\ICSD\ database for new inorganic superconductors.
The model identified about 30 oxides as candidate materials.
Some of these are chemically and structurally similar  to cuprates (even though no explicit structural information was provided during training of the model). Another feature that unites almost all of these materials is the presence of flat or nearly-flat bands just below the energy of the highest occupied electronic state. 

In conclusion, this work demonstrates the important role 
ML models can play in superconductivity research. 
Records collected over several decades in SuperCon and other relevant databases can be consumed by ML models, 
generating insights and promoting better understanding of the connection
between materials' chemistry/structure and superconductivity.
Application of sophisticated ML algorithms has the potential to dramatically accelerate
the search for candidate high-temperature superconductors.

{\small
\section*{Methods}

\noindent \textbf{Superconductivity data.}
The SuperCon database consists of two separate subsets: ``Oxide \& Metallic''
(inorganic materials containing metals, alloys, cuprate high-temperature superconductors, \textit{etc.})
and ``Organic'' (organic superconductors).
Downloading the entire inorganic materials dataset and removing compounds with
incompletely-specified chemical compositions leaves about $22,000$ entries.
In the case of multiple records for the same material,
the reported material's $T_{\mathrm{c}}$'s are averaged, but only if
their standard deviation is less than $5$~K, and discarded otherwise. 
This brings the total down to about $16,400$ compounds,
of which around $4,000$ have no critical temperature reported. Each entry in the set contains fields for the chemical composition, 
$T_{\mathrm{c}}$, structure, and a journal reference to the information source.
Here, structural information is ignored as it is not always available.

There are occasional problems with the validity and consistency of some of the data.
For example, the database includes some reports based on tenuous experimental evidence and
only indirect signatures of superconductivity, as well as reports of inhomogeneous (surface, interfacial)
and nonequilibrium phases.
Even in cases of \textit{bona fide} bulk superconducting phases, important relevant variables
like pressure are not recorded.
Though some of the obviously erroneous records were removed from the data,
these issues were largely ignored
assuming their effect on the entire dataset to be relatively modest. The data cleaning and processing is carried out using the Python Pandas package for data analysis~\cite{Mckinney_Pandas_2012}.

\noindent \textbf{Chemical and structural features.}
The predictors are calculated using the Magpie software \cite{magpie_software}.
It computes a set of 145 attributes
for each material, including:
(\textit{i}) stoichiometric features (depends only on the ratio of elements and
not the specific species);
(\textit{ii}) elemental property statistics: the mean, mean absolute deviation, range, minimum,
maximum, and mode of 22 different elemental properties
(\textit{e.g.}, period/group on the periodic table,
atomic number, atomic radii, melting temperature);
(\textit{iii}) electronic structure attributes: the average
fraction of electrons from the $s$, $p$, $d$ and $f$ valence shells among all
elements present; and
(\textit{iv}) ionic compound features that include whether it is possible to form an ionic
compound assuming all elements exhibit a single oxidation state.

ML models are also constructed with the
superconducting materials in the \AFLOW\ Online Repositories.
\AFLOW\ is a high-throughput \textit{ab initio} framework that manages density functional theory (DFT)
calculations in accordance with the \AFLOW\ Standard~\cite{curtarolo:art104}.
The Standard ensures that the calculations and derived properties are empirical (reproducible), reasonably
well-converged, and above all, consistent (fixed set of parameters), a particularly attractive feature for ML modeling.
Many materials properties important for superconductivity have been calculated within the \AFLOW\ framework,
and are easily accessible through the \AFLOW\ Online Repositories.
The features are built with the following properties:
number of atoms, space group, density, volume, energy per atom, electronic entropy per atom, valence of the cell,
scintillation attenuation length, the ratios of the unit cell's dimensions, and Bader charges and volumes.
For the Bader charges and volumes (vectors), the following statistics
are calculated and incorporated:
the maximum, minimum, average, standard deviation, and range.

\noindent \textbf{Machine learning algorithms.}
Once we have a list of relevant predictors, various ML models can be applied to the
data~\cite{Bishop_ML_2006,Hastie_StatLearn_2001}.
All ML algorithms in this work are
variants of the random forest method~\cite{randomforests}.
It is based on creating a set of individual decision trees (hence the ``forest''),
each built to solve the same classification/regression problem.
The model then combines their results, either by voting or averaging depending on the problem.
The deeper individual tree are, the more complex the relationships the model can learn,
but also the greater the danger of overfitting, \textit{i.e.}, learning
some irrelevant information or just ``noise''.
To make the forest more robust to overfitting, individual trees in the ensemble are
built from samples drawn with replacement (a bootstrap sample) from the training set.
In addition, when splitting a node during the construction of a tree, the model chooses the best split
of the data only considering a random subset of the features.

The random forest models above are developed using scikit-learn --- a powerful and efficient machine
learning Python library \cite{Pedregosa_JMLR_2011}.
Hyperparameters of these models include the number of trees in the forest,
the maximum depth of each tree, the minimum number of samples required to split an internal node,
and the number of features to consider when looking for the best split.
To optimize the classifier and the combined/family-specific regressors, the
GridSearch function in scikit-learn is employed, which generates and compares candidate models from a grid of parameter values.
To reduce computational expense, models are not optimized at each step of the backward feature selection process.
\begin{figure*} 
\centering
\includegraphics[width=\linewidth]{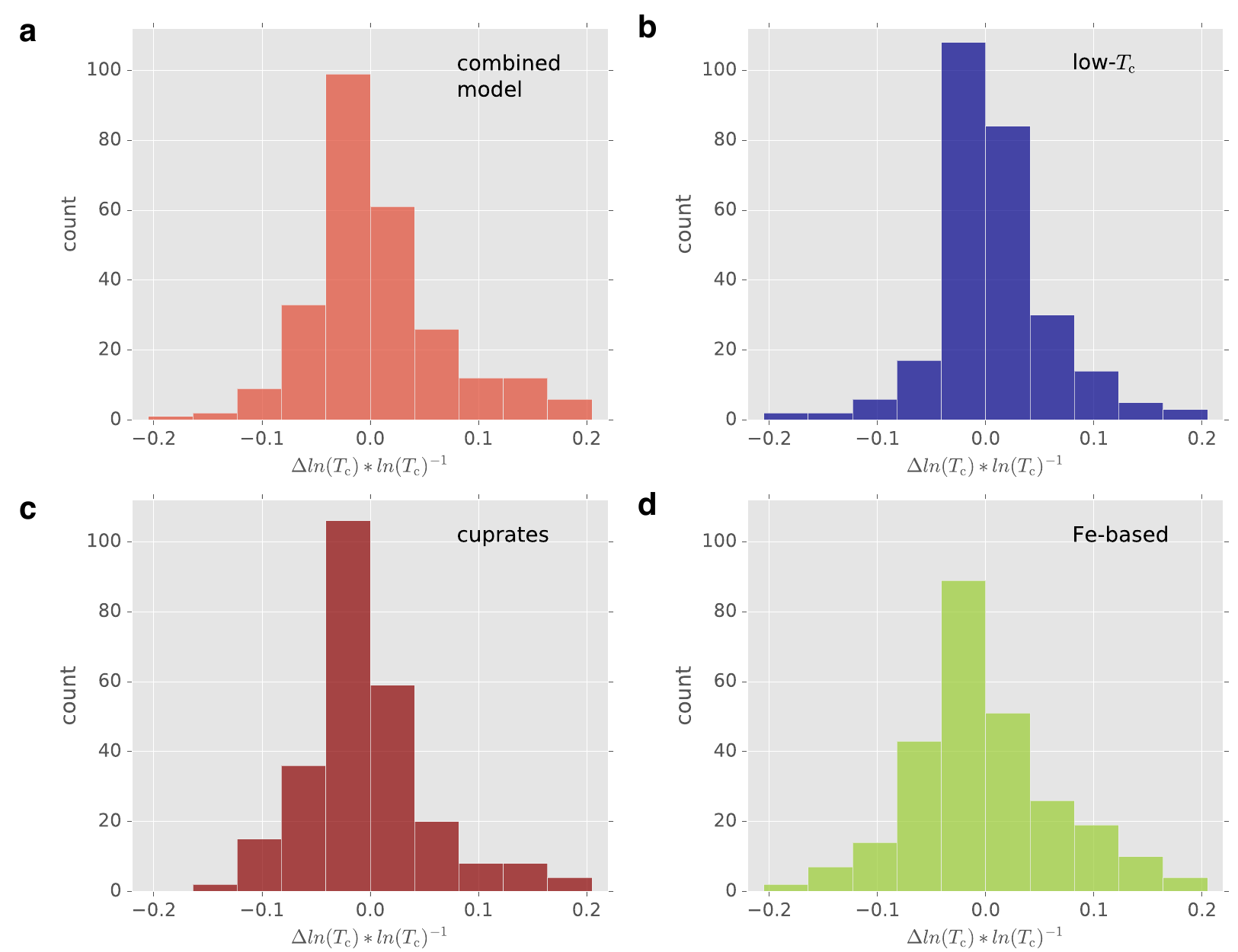}
\caption{\textbf{Histograms of $\Delta\ln(T_{\mathrm{c}}) * \ln(T_{\mathrm{c}})^{-1}$ for the four regression models.}
$\Delta\ln(T_{\mathrm{c}}) \equiv (\ln(T^{\mathrm{meas}}_{\mathrm{c}}) - \ln(T^{\mathrm{pred}}_{\mathrm{c}}))$
and $\ln(T_{\mathrm{c}}) \equiv \ln(T^{\mathrm{meas}}_{\mathrm{c}})$.}
\label{Regr_err}
\end{figure*}

\noindent \textbf{Prediction errors of the regression models.}
Previously, several regression models were described, 
each one designed to predict the critical temperatures of materials from different superconducting groups. 
These models achieved an impressive $R^{2}$ score, demonstrating 
good predictive power for each group. 
However, it is also important to consider the accuracy of the predictions 
for individual compounds (rather than on the aggregate set), 
especially in the context of searching for new materials. 
To do this, we calculate the prediction errors for about 300 materials from a test set. 
Specifically, we consider the difference between the logarithm of the predicted and measured 
critical temperature $[\ln(T^{\mathrm{meas}}_{\mathrm{c}})- \ln(T^{\mathrm{pred}}_{\mathrm{c}})]$
normalized by the value of $\ln(T^{\mathrm{meas}}_{\mathrm{c}})$ 
(normalization compensates the  different $T_{\mathrm{c}}$ ranges of different groups).
The models show comparable spread of errors.
The histograms of errors for the four models 
(combined and three group-specific) are shown in Fig.~\ref{Regr_err}. 
The errors approximately follow a normal distribution, 
centered not at zero but at a small negative value. 
This suggests the models are marginally biased, and on average tend to slightly underestimate $T_{\mathrm{c}}$. 
The variance is comparable for all models, but largest for the model trained 
and tested on iron-based materials, which also shows the smallest $R^2$.
Performance of this model is expected to benefit from a larger training set.
}
\section*{Acknowledgments}
The authors are grateful to Daniel Samarov, Victor Galitski, Cormac Toher, 
Richard L. Greene and Yibin Xu for many useful discussions and suggestions. We acknowledge Stephan R\"{u}hl for \ICSD.
This research is supported by ONR N000141512222, ONR N00014-13-1-0635, and AFOSR No. FA 9550-14-10332.
CO acknowledges support from the National Science Foundation Graduate Research Fellowship under Grant No. DGF1106401. JP acknowledges support from the Gordon and Betty Moore Foundation’s EPiQS
Initiative through Grant No. GBMF4419.
SC acknowledges support by the Alexander von Humboldt-Foundation.


\end{document}